\documentclass{article}

\usepackage{arxiv}

\usepackage[utf8]{inputenc} 
\usepackage[T1]{fontenc}    
\usepackage{hyperref}       
\usepackage{url}            
\usepackage{booktabs}       
\usepackage{amsfonts}       
\usepackage{nicefrac}       
\usepackage{graphicx}
\usepackage{doi}
\usepackage{amsmath}

\title{Extensible Consent Management Architectures for Data Trusts}

\author{Balambiga Ayappane \\
        International Institute of Information Technology, Bangalore\\
	26/C, Electronics City Phase 1\\
	Bangalore, Karnataka, India\\
	\texttt{balambiga.ayappane@iiitb.ac.in}
\And
 Rohith Vaidyanathan \\
        International Institute of Information Technology, Bangalore\\
	26/C, Electronics City Phase 1\\
	Bangalore, Karnataka, India\\	\texttt{rohith.vaidyanathan@iiitb.ac.in}\
 \And
 Srinath Srinivasa \\
        International Institute of Information Technology, Bangalore\\
	26/C, Electronics City Phase 1\\
	Bangalore, Karnataka, India\\
	\texttt{sri@iiitb.ac.in}
\And
 Jayati Deshmukh \\
        International Institute of Information Technology, Bangalore\\
	26/C, Electronics City Phase 1\\
	Bangalore, Karnataka, India\\
	\texttt{jayati.deshmukh@iiitb.org}}

\date{}

\hypersetup{
pdftitle={Extensible Consent Management Architectures for Data Trusts},
pdfsubject={cs.CY},
pdfauthor={Balambiga Ayappane, Rohith Vaidyanathan, Srinath Srinivasa, Jayati Deshmukh},
pdfkeywords={Privacy, Personal information, Legal capacity, Role Tunnel, Data Trusts},
}

\begin{document}
\maketitle

\begin{abstract}
	Sensitive personal information of individuals and non-personal information of organizations or communities often needs to be legitimately exchanged among different stakeholders, to provide services, maintain public health, law and order, and so on. While such exchanges are necessary, they also impose enormous privacy and security challenges. Data protection laws like GDPR for personal data and Indian Non-personal data protection draft  specify conditions and the \textit{legal capacity} in which personal and non-personal information can be solicited and disseminated further. But there is a dearth of formalisms for specifying legal capacities and jurisdictional boundaries, so that open-ended exchange of such data can be implemented. This paper proposes an extensible framework for consent management in Data Trusts in which  data can flow across a network through ``role tunnels'' established based on corresponding legal capacities.
\end{abstract}


\section{Introduction}

A number of services require processing and exchange of private and personal information of individuals. For instance, medical records may need to be exchanged between specialists across hospitals. Similarly, personal information like academic credentials, driving history, credit rating, etc. are routinely exchanged across multiple institutions for providing services. 

Technologies like blockchain provide distributed ledgers and audit trails, that protects the integrity of the information exchange. However there is still a need for formalisms for encoding and enforcing the \textit{policy} governing the exchange of sensitive data. 

Current day inter-organizational information exchange are usually modelled in the form of web services, that implement authentication and access-control mechanisms to regulate the exchange~\cite{borger2008method,schewe2005conceptual}. Here, specific applications are granted access by having them register with the web service, and providing them with an application id and access keys. In such mechanisms, privileges are tightly connected to the \textit{identity} of the client application and its owner. This makes it difficult to seamlessly extend access rights to other legitimate users. For instance, if a person who is handling sensitive data through their authorized application is incapacitated or deceased, their successor cannot seamlessly take on their role unless they have the identity-based access credentials. 

Research literature has long since advocated RBAC or Role-based access control mechanisms to greatly simplify specification of access privilege policies~\cite{bertino2001trbac,sandhu2000nist,sandhu1996role}. A role represents a competency to do a particular operation, and it connects a set of people or applications, with a set of privileges. Access privileges are associated with roles, rather than with individuals, and the association of individuals with roles are dynamic. Authentication now involves not only proving one's identity, but also proving one's role. 

RBAC models are typically implemented within an organizational context. This means that the RBAC mechanism is situated within a larger semantic framework that establishes associations of users with roles. RBAC frameworks are also extended for inter-organizational workflows~\cite{fabian2015collaborative,kang2001access,park2017role}. Several approaches are adopted for extending an RBAC framework across organizations. These include creation of ``virtual organizations'' representing role granting authorities for inter-organizational interactions and/or mapping of roles across organizations. 

Before the existence of banks, people handled their money and finances themselves. These financial details were private and not known by others. The state of financial affairs depended on how well people could manage their finances. However, once banks came up, it handled most of these aspects of financial management. A bank became a trusted place where people could deposit their money. Financial details were now available with the person and the bank. An added benefit of a bank is that the money is not just lying stagnant in a bank but gradually increasing based on the interest rates. As well as the collective financial assets are not just useful for the person or for the bank but are beneficial for the economy as a whole.

Similarly, so far either we manage our own data or it is in complete control of other agencies. Managing our data gives us full control on the data however it requires specific knowledge and skills of data management and also even in an anonymized version, it cannot be used by others. On the other hand, we lose control once the data is handed over to other agencies which manage and share our data. This calls for a data trust like a bank, which is a trusted agency to manage the data. It manages all the individual and organization level data and can also share it to other trusted agencies after processing like anonymization and aggregation. Thus data is not just relevant for the individual or organization to which it belongs but to other trusted agencies. We discuss further in detail how our proposed Multiverse framework can be used to design and build data flows in data trusts.

More recently, there has been increased interest in open-ended data dissemination in the form of ``open data'' for greater common good. With increasing numbers of governance and administrative workflows appearing online, there is also an increased need for exchanging data across several entities with little or no inter-organizational authorities for managing the integrity of data exchange. 

While open data improves transparency and accountability in public workflows, it also brings with it challenges of privacy and security leading to several contradictory requirements~\cite{agrawal2014integrity,eckartz2014decision,meijer2013bridging,meijer2014reconciling,srinivasa2014characterizing}. Specifically, open-ended data exchange is characterized by three divergent concerns~\cite{srinivasa2014characterizing}: \textit{transparency}, \textit{privacy}, and \textit{security}. Transparency requires relevant data to be shared publicly in order to uphold integrity of a public action. Privacy on the other hand, requires data to be withheld in order to protect the dignity and liberty of individuals. Security pertains to collective good, where open-ended sharing of certain sensitive information, can endanger a community or country. 

In addition to the above concerns, there is also a need for \textit{legitimate} open-ended sharing of private and/or sensitive information in times of crisis, to protect public health and order. For instance, public health management in the time of Covid crisis requires private and sensitive information about patients suffering from Covid to be shared with several stakeholders like doctors, administrators, volunteer organizations, etc. 

In such cases, there is no overarching virtual organizational structure, or role granting and mapping authority. The number of disparate entities requiring the data may keep changing over time, and may not be known a priori. This makes it infeasible to apply existing approaches to inter-organizational privilege management. 

In this paper, we propose a modular, extensible framework called ``Multiverse'' to manage legitimate exchanges of sensitive data in an open-ended fashion, without the need for an overarching organizational framework to enforce the integrity of data exchange. In Section \ref{sec:RelatedWork} we discuss some of the existing models of access control systems. Details of the ``Multiverse'' framework are presented in Section \ref{sec:multiverse}. Section \ref{Sec:adverserial} discusses a variety of adversarial scenarios and how it can be handled by the ``Multiverse'' framework and Section \ref{sec:caseStudies} presents a couple of case studies where it is useful. Conclusions and future directions are presented in Section \ref{sec:conclusions}. 

\section{Related Work}\label{sec:RelatedWork}

Access control systems act a mediator between users and data/ resources to grant or deny access based on the underlying security policy~\cite{samarati2000access}. Access control systems can be broadly classified into two categories: encryption based systems and proof based systems~\cite{hengartner2005exploiting}. Encryption based systems encrypt the data and send it off to the individual. The individual needs to have the appropriate key in order to decrypt the data. On the other hand, proof based systems require the individual to produce all the necessary proofs required to authenticate their identity and after authentication, the data is shared with the individual. It is difficult to provide fine grained access control using encryption based methods without increasing the number of keys as well as it is computationally expensive to manage these systems specially at a large scale. Proof based methods and its variants on the other hand, can better handle fine-grained granularity and constraints.

Individuals can be granted or revoked access based on their identity, which is known as Identity-Based Access Control (IBAC)~\cite{kunzinger2006integrated,hengartner2005exploiting}. In these designs, the individuals need to prove their identity using authentication techniques like the use of passwords, biometrics, or  combinations of public and private keys. Once an individual's identity is authenticated, they can access the required data. However, in large organizations and teams spanning multiple organizations, it is difficult and cumbersome to manage access controls of all the stakeholders individually in this manner.

Role-based access control (RBAC) methods~\cite{bertino2001trbac,sandhu1996role} were designed so that permissions can be granted to users based on their roles rather than their identity. This access control design is more effective as changing roles of an individual automatically updates their privileges. The roles can be assigned to the individuals by an authorized individual. A RBAC policy is designed using role-permission, user-role and role-role relationships. There are variants of RBAC models like models which can handle role hierarchies, constraints, triggers and temporal dependencies, teams within the organization~\cite{sandhu1996role,bertino2001trbac,thomas1997team}.


Using open data has its own benefits as well as challenges~\cite{janssen2012benefits}, however it is difficult to make appropriate use of open data in the absence of open data management systems which can handle handle large volume of diverse data such that it can be securely accessed by different types of users. There are a few open data management systems~\cite{roman2018datagraft,da2014dendro} however most of these systems focus on data cleaning and pre-processing so that it can be stored in a database or graph etc. To the best of our knowledge, there does not exist a system which combine data pre-processing, data storage, data retrieval, data visualization along with secure access control system of data specifically for open data systems. 

Data containing personally identifiable information (PII) needs to be handled much more carefully than say population level aggregate data, since it can reveal the identity of individuals and in turn put them at risk. For example, in healthcare setting even anonymized data can be used to infer patient's identity based on their diagnosis details, location etc.~\cite{malin2007computational,loukides2010disclosure,malin2011identifiability}. There are encryption based and data masking techniques to manage access to personal and personally identifiable data. However, there is a dearth of computational models which can define access mechanisms for data aligned to the laws of that region or country. It is especially crucial with data protection laws like EU's General Data Protection Regulation (GDPR)~\cite{EuropeanParliament2016a}, Sweden's  Data Act, Philippines's The Data Privacy Act, India's Digital Personal Data Protection Act, 2023~\cite{dpdpa}, California's Consumer Privacy Act etc. being defined around the world. 

Such privacy preserving frameworks have also been proposed for Non-Personal Data. According to the Indian Non Personal Data Protection Draft~\cite{NPD}, any data that cannot be used to identify an individual is called Non-Personal Data. Such data can be collected from communities, organizations and even the general public. This data can be processed to derive insights that are beneficial socially and economically. This draft proposes roles like community which are owners of the non-personal data, data custodians who collect data from communities, data trustees which create High-Valued datasets with socio-economic value and an NPDA authority to adjudicate in case of disputes.

Another important aspect of legitimate exchange of data, is \textit{consent}. In recent times many data protection regulations like General Data Protection Regulation (GDPR) of EU, Digital Personal Data Protection Act, 2023 (DPDPA-2023) of India etc. have emphasized on the need to collect, share and use personal data according the Data Principal's consent. The Indian Non Personal Data Protection Bill Draft proposes the same for Non-Personal Data collected from communities or organizations.

In the context of personal data, DEPA\footnote{Data Empowerment and Protection architecture https://www.niti.gov.in/sites/default/files/2020-09/DEPA-Book.pdf} proposes a national standard for managing consent in India using 'ORGANS' framework. The ORGANS framework stands for Open Standards, Revocable, Granular, Auditable, Notice by Design and Secure by Design. 

Many current consent solutions use blockchain~\cite{albanese2020dynamic,genestier2017blockchain,xu2021ac2m,rantos2019advocate} for ledgering and auditing. Most existing consent management solutions are designed for specific  domains ~\cite{shrivastava2023comprehensive,xu2021ac2m} like healthcare, finance etc.

The common approach to obtaining consent has been based on the Autonomous Authorization model~\cite{schermer2014crisis} but this approach may not be suitable for managing consent of collectively-owned community or organization data.

\section{The Multiverse Framework}\label{sec:multiverse} 

The proposed framework called ``Multiverse'' to create an extensible, open-ended infrastructure for consent management and legitimate exchange of data, is described in this section. A Multiverse framework, also called a ``Frame'' $F$ is made up of the following building blocks: 

\begin{equation}
    F = (W, D, A, T) 
    \label{eq:frame}
\end{equation}

Here $W$ is a set of containers called ``worlds'' that represent the semantic boundary or legal jurisdiction within which, a data element is accessed and processed. $D$ represents the set of all data elements or ``resources'' that are being shared. The term $A$ represents ``agents'' which could be users or application programs that produce and consume resources. The term $T$ represents a set of ``templates'' where each template defines a set of access points through which data may be accessed, and a set of relationship types with which worlds can be related. 

A world represents the basic unit within which data is accessed. Every agent $a \in A$ has its own corresponding world named as $world(a)$. In addition to representing agents, a world could represent any semantic entity or legal jurisdiction. Some examples of worlds include: institutions, town municipalities, undivided families, resident welfare societies of communities, etc. Every data element is published within the boundaries of a world, and data is only exchanged between worlds. An agent $a$ only publishes its data into $world(a)$, and also reads data only from $world(a)$. 

\begin{figure*}
    \centering
    \includegraphics[width=3.5in]{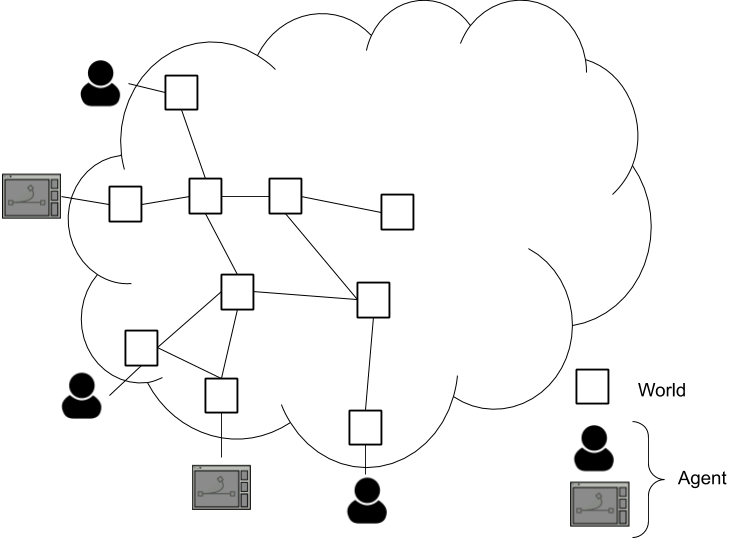}
    \caption{Multiverse framework}
    \label{fig:multiverse}
\end{figure*}

Figure~\ref{fig:multiverse} depicts a multiverse schematically. The multiverse is a network of worlds connected by one or more relations defined in templates. Data are published within worlds and exchanged between them based on a system of \textit{legal capacities} explained later on. Agents, which include users and application programs, lie outside of the multiverse cloud, but have a representation for themselves in the form of a semantic world, within the multiverse. 

Worlds could be \textit{contained} within one another. If world $w_2$ is contained within world $w_1$, this is represented as $w_2 \triangleright w_1$. Containment of a world is called its ``jurisdictional location'' or simply ``location'' that represents a system of privilege inheritance explained later on. 

Each world may implement one or more \textit{templates} $t \in T$, that gives it a semantic characterization in the form of a set of data access points and relationship types with other worlds. Any template $t \in T$ is made up of the following elements: 

\begin{equation}
    t = (Dat, Rel) 
    \label{eq:template}
\end{equation}

Here $Dat$ represents a set of \textit{data access points}, and $Rel$ represents a set of \textit{relationship specifications}. A data access point represents a gated interface through which a given data element may be accessed. Any data access point $dat \in Dat$ comprises of the following elements: 

\begin{equation}
    dat = (Q, C, P) 
    \label{eq:dat}
\end{equation}

Here $Q$ is the query with which the data element is accessed. The terms $C$ and $P$ represent \textit{legal capacity} and \textit{purpose code} respectively, which are both explained later. 

The relationship specifications $Rel$ specifies the kind of relationships that the world can implement with other worlds, as well as the kinds of relationships that the world can accept from other worlds. 

A relationship specification may be of two kinds-- an \textit{incoming} relationship specification, and an \textit{outgoing} relationship specification. These are represented as $Rel_i$ and $Rel_o$ respectively. A relationship specification comprises of the following elements: 

\begin{gather}
\begin{split}
rel_i = (name, constraints, \\ privileges, purposes)\label{eq:reli}
\end{split}
\\
rel_o = (name, constraints, roles)\label{eq:relo}
\end{gather}


A relationship has a name at its incoming end and its outgoing end. The incoming relationship name is also called a \textit{role}. Any agent entering a world through a relationship, where the incoming name of the relationship is $r$, is said to be playing the role $r$ in the world. 

\begin{table*}[]
\centering
\begin{tabular}{|p{1.3in}|p{1in}|p{2.8in}|}
\hline
\textbf{Constraint} & \textbf{Specification} & \textbf{Meaning} \\
\hline
\hline 
Template & implements(t) & The source or target world needs to be implementing template $t$ for the relationship to be valid.\\
\hline 
Template Relationship & relt(name, t) & The source or target world needs to have a relationship named $name$ with a world implementing template $t$.\\
\hline
Identity Relationship & relid(name, id) & The source or target world needs to have a relationship named $name$ with world identified by $id$.\\
\hline
\end{tabular}
\caption{Relationship constraints}
\label{tab:rel} 
\end{table*}

Both outgoing and incoming relationships are subject to a set of constraints. Table~\ref{tab:rel} specifies different kinds of constraints on a relationship. The template constraint: \[implements(t)\] in an outgoing relationship specification means that the target world with which this relationship is being established, should be implementing template $t$. Similarly, such a constraint for an incoming relation means that, the relationship can be accepted only if the recipient world is implementing template $t$. Here, the reference to template $t$ is in the form of a universally uniquer ID like a URI. 

Hence for example, in a template called $Person$, we can specify that an outgoing relationship called $Employee$ can be established with a target world, only if the world implements a template called $Company$. Similarly, for a template called $Company$ there can be an incoming relationship called $Employee$, which may have a constraint that the source world should have implemented a template called $Person$.

The $relt(name,t)$ constraint specifies that the target or source world should have a relationship named $name$ with a world that implements template $t$. 

Hence for example, in the template specification of a $Person$, we can specify a relationship called $Patient$ with another world $w$, only if $w$ has a relationship named $Doctor$ with a world that is implementing a template called $Hospital$. In other words, a person can be related to another person as a patient, only if the other person is a doctor at some hospital. 

In addition to template and relationship specifications, a constraint could also identify specific worlds with their unique identifiers, using the $relid(name, id)$ specification. 

Any given $rel_i$ or $rel_o$ specification may have multiple constraints specified. In such cases, all the specified constraints need to be satisfied, for an instance of the relationship to be formed. 

The $privileges$ part of the relationship specification for $rel_i$, represents the privileges that any agent obtains, when traversing a relationship edge. 

\begin{table*}[]
\centering
\begin{tabular}{|p{1in}|p{1in}|p{2.7in}|}
\hline 
\textbf{Privilege class} & \textbf{Privilege} & \textbf{Interpretation}\\
\hline
\hline 
Resource & read & all forms of read queries on a resource \\
    & write & write or modify a resource\\
    & delete & delete a resource\\
    & template & access templates visible to this world\\
\hline
World & edit & modify privileges on the current world, including management of templates and deleting the world\\
    & relocate & move the world from its current location to another location \\
    & create & privilege to create worlds inside this world \\
\hline 
\end{tabular}
\caption{Role privileges} 
\label{tab:privileges} 
\end{table*}

\noindent An incoming agent who enters a world through a relationship, gets the role specified in $rel_i$, and the corresponding privileges associated with it. Table~\ref{tab:privileges} details a set of privilege classes, that apply respectively to a set of operations over resources (including templates), and the world itself. A role having a \texttt{resource.read} privilege for example, enables the role player to read resources hosted by this world. 

The $purposes$ element of $rel_i$ specifies a set of legitimate reasons or ``purpose codes'' for which a particular activity needs to be performed. Annotating a purpose code for each data access, helps in establishing official legitimacy for the access. The $purpose$ code is represented as an enumerated list of values. 

The $roles$ element of $rel_o$ in Eq~\ref{eq:relo} specifies the roles within the source world that are entitled to traverse the given relationship edge. Hence, in a given world $w$, an outgoing relationship specification of the form: $(r_o, c, p)$ represents that any agent playing the role $p$ can traverse the relationship edge $r_o$ to act as a representative of the source world $w$, in the target world. 

Every world also has a role called $owner$, which trivially has all privileges. The creator of a world is its default owner, but may add other owners and/or give up the owner role to other agents. 

\begin{figure*}
    \centering
    \includegraphics[width=3.5in]{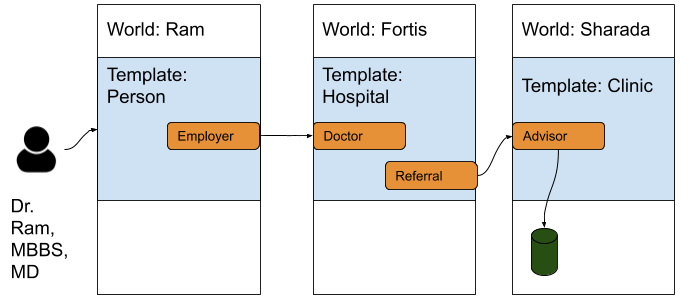}
    \caption{Role Tunneling}
    \label{fig:roletunnel}
\end{figure*}

When an agent traverses a relationship to reach a new world, the \textit{legal capacity} in which the agent performs any operation in the target world is a concatenation of all the roles played by the agent beginning from the world representing the agent. 

Figure~\ref{fig:roletunnel} shows an example. Here, an agent who is a user named Dr. Ram is accessing some data stored in a world called Sharada. Sharada has implemented a template called Clinic, and it is in relationship with another world called Fortis, which has implemented a template called Hospital. The world for the user Ram, is also in relationship with Fortis, with the role of Doctor. The relationship between the Hospital and the Clinic enables a Doctor of the Hospital to appear as Advisor in the Clinic, which gives them some privileges over the data. 

Here, when Dr. Ram accesses some data element $d$ stored at Sharada, the data access point would look as follows: 


\begin{eqnarray*}
    dat & = & (read(d),\\ 
    & & Advisor(Sharada): \\
    & & Doctor(Fortis):Owner(Ram),\\
    & & Diagnostics)  
\end{eqnarray*}

The last term $Diagnostics$ represents the purpose code, indicating the official purpose for which the data is being accessed. The string: $Advisor(Sharada):Doctor(Fortis):Owner(Ram)$ represents the \textit{legal capacity} in which the access is being made. This represents a string of role and world specifications that leads up from the agent to the data source. 

The string representing the legal capacity is called a \textit{Role Tunnel}, since it creates a legal pathway from the agent to the dataset based on legal arrangements between worlds. 

A role tunnel is valid if each element in the role tunnel satisfies their corresponding $rel_i$ constraints, and the last element in the tunnel represents the $Owner(p)$ role, where $p$ is the id of the agent performing the access. 

Each resource stored in a world also has stored along with it, the legal capacity by which it was brought there. Formally, a resource $s$ in a world has the following fields: 

\begin{equation}
    s = (d, c, ttl) 
\end{equation}

Here, $d$ is the data element, and $c$ is the legal capacity by which the data element came to be stored in the world. If the data element is local to the world and was not imported from elsewhere, the $c$ field would be null.  

A data element with a string of multiple roles for its legal capacity represents a \textit{remote} data element brought in from a remote source. All remote data elements also have a ``Time To Live (TTL)'' parameter, which indicates the length of time until which it can be stored at the remote location. After the TTL expires, the data needs to be fetched again through a legal role tunnel. 

Every access of a data element involves checking the validity of the legal capacity. A role tunnel of the form: $r_n(w_n):\dots:r_2(w_2):r_1(w_1):Owner(w)$ requires $n+1$ integrity checks before the data access can be made possible. If the legal capacity of the agent fails to hold when accessing a remote data element that is cached in its world, then the data element is removed from the world. Subsequent access to the data element requires the agent to approach the source world of the data element through an legal role tunnel, and fetch it once again. 

Note that a legal capacity represents a \textit{logical} tunnel. A role tunnel of the form $r_n(w_n):\dots:r_2(w_2):r_1(w_1):Owner(w)$ does not require the data to \textit{physically} flow through all the intermediate worlds in the tunnel between $w_n$ and $w$. The interim worlds are required only for establishing the legitimacy of the data access. The interim worlds should be reachable and be able to validate the given role at the time of access. 

Data and network level security in the form of encryption and secure communication, will need to be implemented in addition to the mechanisms of the Multiverse. The Multiverse framework only provides a system for creating legally tractable privilege frameworks across independent institutional contexts. 

\paragraph*{Role inheritance:}

Containment of worlds have special semantics in terms of inheritance of roles. Suppose world $w_2$ is contained in $w_1$ and both implement a template $t$. In such cases, any agent playing a given role $r$ in the container world, also gets the privilege of role $r$ in the contained world. This enables aggregation of similar worlds into a larger world, and defining privileges on the larger, container world, rather than on each world individually. 

Hence for example, if a Hospital $H$ has several branches each implementing a template of type Hospital, with each branch contained within the larger world $H$, then any agent playing a role (say, $Doctor$) in $H$ would also get to play the same role with the same privileges, in all branches contained in $H$. 


\paragraph*{Template visibility:} Templates are treated like any resource, and can be created within any world by agents who have write privileges on the world. Other worlds that have read privileges on a given world $w$ can access and implement the templates defined in world $w$. 
When a template $t$ that is defined in world $w$ is used in another world $w'$, it is treated as a remote resource in $w'$ and the role tunnel with which $t$ was accessed, is stored along with $t$, in addition to the TTL parameter. Use of the template data access points, or creation or deletion of relationship instances of the template will require the legal capacity of the template to be satisfied.

For instance, let template $t$ in world $w$ be accessed through a role tunnel $r_2(w_2):r_1(w_1):Owner(w)$. The use of this template for accessing a data element and/or defining a relationship, will require the above legal capacity to be valid. The template $t$ will also need to be retrieved once again after its $ttl$ has expired. An expired template will return false for all its relationships and data access points. At any point during the use of a template, if the template role tunnel is not satisfied, the template is marked as expired and will be unusable, until it is retrieved again from the source. 

Templates can also be subclassed from other templates to form a conceptual subsumption tree. If template $t'$ is a subclass of template $t$, then $t'$ inherits all the data access points and relation specifications from $t$. The subclass $t'$ can override definitions of data access points and/or relation specifications to apply to the world implementing the subclass template. 

\paragraph*{Access risk:} Suppose that a remote data element $d$ is cached in a world $w$ using the following role tunnel: $r_n(w_n):\dots:r_1(w_1):Owner(w)$. Accessing this data element $d$ requires $n+1$ integrity checks to be made. Now suppose that a given role $r_i(w_i)$ is implemented by world $w_i$ using template $t_i$ that itself is fetched using yet another role tunnel $rt(t_i)$. Validating  $r_i(w_i)$ will now require validating the role tunnel for the template that has defined $r_i$. This validation may in turn require further validations of further templates along the way. 

In order to reduce and limit this unfolding of role tunnel integrity checks, data access is characterized by a notion of \textit{access risk}, denoted by a parameter $\rho \in [0,1]$. This represents a decay parameter computing a probability function, which defines whether an integrity check is made at a given level. 

For the initial level of data access (also called level 0), where integrity check is done for the role tunnel from which a data element is retrieved, the integrity check is performed with a probability $(1-\rho)^0$. For the next level of integrity checks, where the templates defining the roles are themselves validated, integrity check is initiated with a probability $(1-\rho)^1$. Similarly, integrity check at level $k$ is initiated with a probability $(1-\rho)^k$. Hence, the higher the value of $\rho$ the lesser the levels to which integrity check is performed, and the greater the access risk. 

Access risk is a parameter that is set by the agent performing a read operation, balancing between speed of access and guarantee of legal authenticity of the access. 

\section{Adverserial Scenarios}\label{Sec:adverserial} 

One of the ways in which the proposed Multiverse framework differs from Roles Based Access Control (RBAC) is that it can be used for applications like data trusts where there is an absence of an overarching role-granting authority. Role specifications are defined in templates that are in turn defined within worlds and exchanged across them based on access privileges. 

While this provides enormous flexibility and scalability for the access control framework, it also opens up questions about how easy would it be for the framework to be compromised. 

In this section, we will consider several adverserial scenarios that could potentially affect the integrity of data exchange, and see how the framework addresses such situations. 

\paragraph*{Scenario 1: False implementation of a template:}

One of the constraints for a world to form a relationship with another world, is the $implements(t)$ that requires the source or target world to have implemented template $t$. Since any world can implement a given template, it can be possible that the implementing world is a bogus world that appears like an instance of $t$. 

For instance, suppose a role of type $Doctor$ can be established between a person and a world of type ``Hospital'' (that is, the world has implemented the ``Hospital'' template). Since any world can implement any template, it could be possible that the world is not actually a hospital, but a bogus world implementing the template. 

Such a scenario is possible, only if the ``Hospital'' template is publicly available. To prevent fake representations, important templates should be defined in a world representing a certifying authority, and read access granted to worlds based on an offline verification of their authenticity. 

\paragraph*{Scenario 2: False implementation of a relationship:} The $relt(name, t)$ constraint for a relationship, require the source or target world to have a relationship called $name$ with a world implementing template $t$. 

In such a case, there could be two levels at which information can be falsified-- either template $t$ does not have a relationship named $name$, and/or the world implementing template $t$ is a bogus world. 

In either case, the main security mechanism is to control the definition of $t$. If template $t$ is defined by a certified authority and be made accessible to worlds based on offline validation of their credentials (which is a one-time activity), both levels of falsification can be addressed. 

\paragraph*{Scenario 3: Unauthorized read of third-party data from a world:}

Suppose that Dr. Ram has accessed data about a patient from Sharada clinic from the example from Figure~\ref{fig:roletunnel}. When the resource is copied to the world of Dr. Ram, would it now be accessible to other agents who have a read privilege on this world? 

To answer this, we need to note that the \textit{legal capacity} with which the data element was accessed is also stored along with the data element. In this example, the legal capacity is: $Advisor(Sharada):Doctor(Fortis):Owner(Ram)$. This Role Tunnel representing the legal capacity is stored along with the resource in the $Ram$ world, and has to be valid at the time of accessing the data element. Hence, another agent who is trying to access this data element, will be able to do so, only if the agent satisfies all the roles in the Role Tunnel: $Advisor(Sharada)$, $Doctor(Fortis)$ and $Owner(Ram)$. That is, the agent should not only be listed as a co-owner of the world $Ram$, but should also be listed as a $Doctor$ in the world $Fortis$ and as $Advisor$ in the world $Sharada$. 

\paragraph*{Scenario 4: Malicious Representation:} 

In the example from Figure~\ref{fig:roletunnel}, suppose that the hospital has implemented its $Hospital$ template by downloading it from a regulatory agency $R$, that recognizes hospitals and issues certificates and templates for their operations. Suppose that the hospital loses its recognition due to some malpractice, and is no longer eligible to use the $Hospital$ template. However, the hospital continues to implement the template, even though it was legally required to discontinue its use. What would be the repercussions of such a case? 

There are two safeguards that addresses cases involving such malicious intermediaries. The first is the $ttl$ parameter for the template, which limits the duration until which, the template will be illegally valid. The second safeguard is the access risk $\rho$ parameter by the agent performing a read. If the data being accessed is very sensitive, the reader may set the access risk $\rho$ to a low value, which will force integrity check for the template that defines a role. 

\section{Case Studies}\label{sec:caseStudies}

In this section, we will consider some case study applications where a Multiverse framework would be useful. 

\subsection{CET Score Verification}\label{subsec:caseStudyCET}

Many countries have some form of a Common Entrance Test (CET) for graduate admissions. Applicants who take the test use these scores to gain admissions in universities. The number of universities who recognize CET scores may be large, and may vary over time. In addition, several other organizations may also consider CET scores for hiring employees. 

These organizations will need to independently verify scores of an applicant from the CET database. Considering these scores are shared using a data trust, the verification process can be  securely automated using the Multiverse framework as follows:

\begin{figure*}
    \centering
    \includegraphics[width=3.5in]{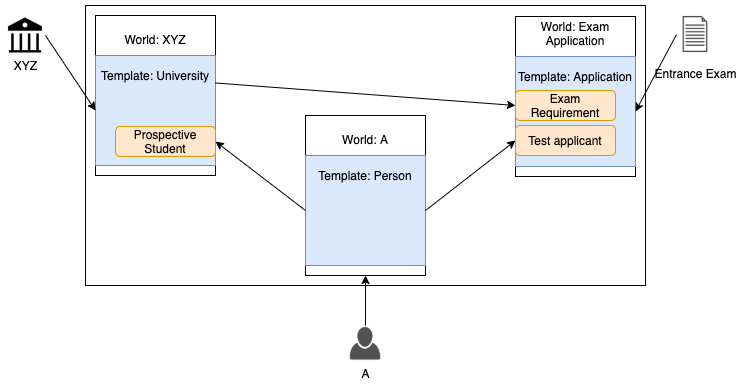}
    \caption{Common Entrance Exam Score Verification Application}
    \label{fig:cet}
\end{figure*}

Applicant $A$ takes the $CET$ which is required for admission at $XYZ$ University. In this setup as shown in Figure \ref{fig:cet}, there are three entities student $A$, university $XYZ$ and $CET$ administering organization, each of which have their own worlds. Applicant $A$ implements the template of $Person$, $CET$ implements the template of $Application$ and University $XYZ$ implements the template of $University$. Following relationships exist among the worlds: Applicant $A$ plays the role of \textit{prospective student} in the world of $XYZ$ University and a \textit{test applicant} for $CET$ application. Once $A$ has completed the $CET$, the scores are stored in their database. $A$ also informs the $CET$ application regarding universities / organizations that s/he is applying to. In turn, $XYZ$ university requests to access the $CET$ scores of applicant $A$ and if the constraints such as applicant $A$ exists, has valid scores, and has applied to university $XYZ$ are all satisfied, then $A's$ scores are securely shared with $XYZ$ university.

\subsection{Identity Validation}\label{subsec:caseStudyUID}

Most countries have a unique ID (UID) of all its citizens. It is used to uniquely identify the citizens and this UID is mandatory for a variety of purposes like opening a bank account, buying / renting a property etc. Even in this scenario, Multiverse framework can be used as follows: 

\begin{figure*}
    \centering
    \includegraphics[width=3.5in]{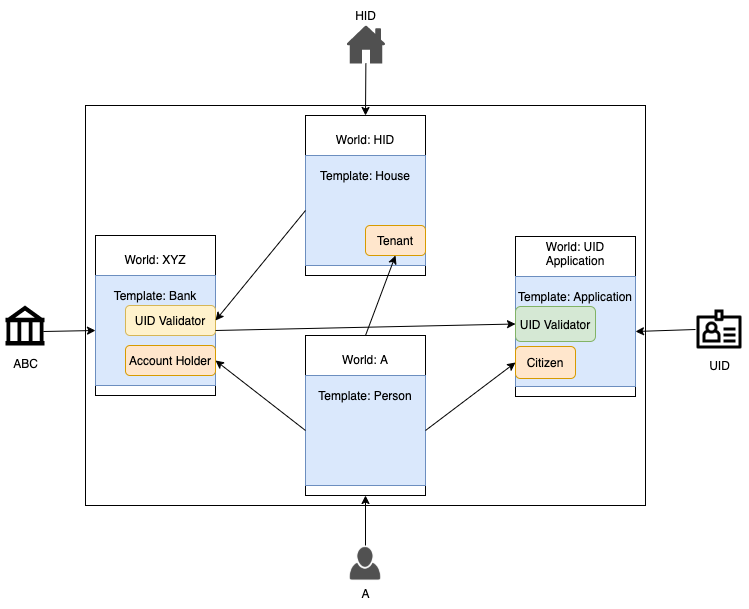}
    \caption{UID Validation Application}
    \label{fig:uid}
\end{figure*}

As shown in Figure \ref{fig:uid}, let's say person $A$ is a citizen and his UID details are stored in the UID application. When $A$ wants to open an account in $ABC$ bank, the bank validates the UID details from the UID application. Since the bank is a well known entity which has been pre-verified and pre-approved, it has read rights on the UID world. Next, $A$ wants to rent a house $HID$. Before $A$ is accepted as a tenant, house $HID$ needs to verify the identify of $A$. Since house $HID$ is not a centralized entity, it does not have direct access to the UID application. However, it is a well-known fact that $A's$ identity is valid if s/he has an account in $ABC$ bank. And thus house $HID$ accepts bank account details  (like account number and address) as valid identity proof of $A$. The role tunnel is complete if $A$ requests $ABC$ bank to share the account details with $HID$.  

\subsection{Data Trust}
\begin{figure*}
    \centering
    \includegraphics[width=5.5in]{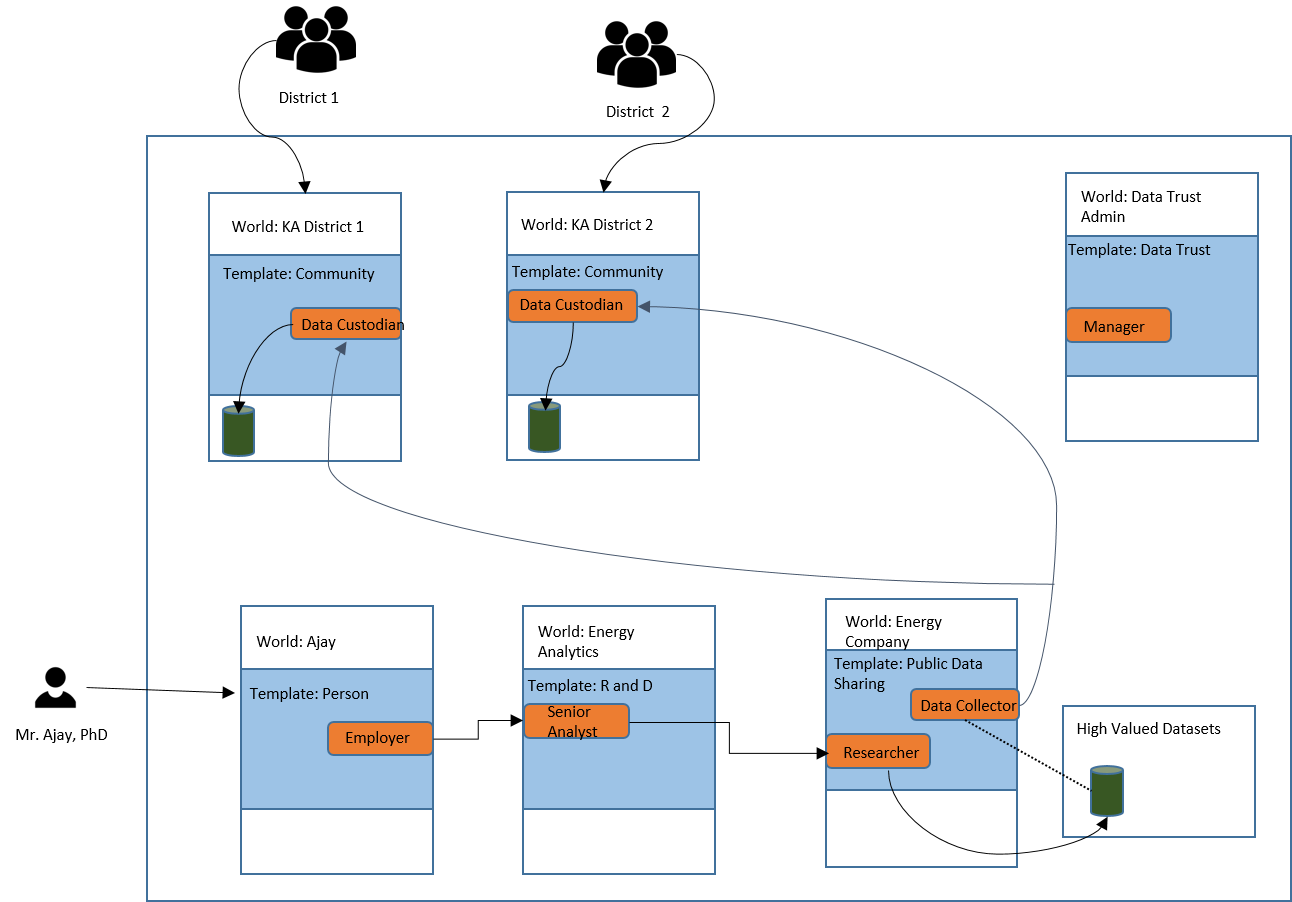}
    \caption{Multiverse Framework for a Data Trust}
    \label{fig:roletunnel_npd}
\end{figure*}

Data Trusts enable legitimate sharing of data between Data Provider (who may be data subjects or data custodians) and Data Requester. Data custodians may collect data about communities and can leverage these platforms to share anonymized version of this data with data requesters or contribute to the creation of 'High Valued Datasets'~\cite{NPD}. For example, A data custodian say an Energy Company has data about electricity usage of a community. This anonymized dataset will have data about the community who have subscribed to this service. This energy company places such data in a data trust which is used to create a 'High-Valued Dataset'. Now, an organisation interested in this High Valued Dataset requests the Data Trust for access to this data. Consent of the community is derived from and managed according to the community's data sharing policy. Figure~\ref{fig:roletunnel_npd} shows how Multiverse works to achieve such data sharing in a Data Trust.


 Here, the agent is represented by a user named Mr. Ajay  who is a Senior Analyst with  Energy Analytics. He needs to access some data stored in a world called Energy Company. Energy Company has implemented a template called Public Data Sharing,  and it is in relationship with Energy Analytics, which has implemented a template called R\&D. The world for the user Ajay, is also in relationship with Energy Analytics, through the role of Senior Analyst. The relationship between the R\&D and the Public Data Sharing enables a Senior Analyst of the R\&D to appear as Researcher in the Public Data Sharing template, which gives him some privileges over the 'High-Valued dataset'. 

Here, when Ajay accesses some data element $d$ stored at Energy Company, the data access point would look as follows: 


\begin{eqnarray*}
    dat & = & (read(d),\\ 
    & & Researcher(Energy Company): \\
    & & Senior Analyst(Energy Analytics):Owner(Ajay),\\
    & & Research)  
\end{eqnarray*}

\section{Conclusions}\label{sec:conclusions} 

Data utility needs to contend with three conflicting concerns-- transparency, privacy and security. Most of the solutions to address these concerns have thus far required a larger institutional framework, that regulates access. Extending the legitimacy of access control across organizational boundaries in an open-ended fashion had always been a challenge. 

The Multiverse framework proposed in this paper addresses this problem, and uses role tunneling as the mechanism for extending inter-organizational access regulations in an open-ended fashion. The Multiverse framework only addresses legitimacy of access control. Protection of the data itself is a different issue that is addressed by encryption and secure communication protocols. Similarly, protection of the data after it has been accessed-- for example, by malicious agents taking a photograph of the data displayed on their screens-- are also outside the scope of the framework. 

The Multiverse framework is primarily meant to record and establish legal channels for handling of sensitive data, and for establishing provenance and audit logging of data access in the form of role tunneling.

\bibliographystyle{plain}
\bibliography{arxiv}
\end{document}